\documentclass[12pt]{article}
\def\ii{\'{\i}}

\def\be{\begin{equation}}
\def\ee{\end{equation}}
\def\ba{\begin{eqnarray}}
\def\ea{\end{eqnarray}}

\def\fr{\frac}

\def\iv{\frac{1}}

\newcommand{\bx}
 {\hbox to.77778em{\hfil\vrule\vbox to.675em
 {\hrule width.6em\vfil\hrule}\vrule\hfil}}
\def\rmd{\mathrm{d}}

\def\ni{\noindent}

\def\vs{\vspace{0.5cm}}
\def\hs{\hspace{0.5cm}}

\usepackage{epsfig}

\begin{document}
%\thispagestyle{empty}
%{\tiny Lambda.tex 20010716}
%\vs

\centerline{\huge \bf Cosmological Constant from}
\vs

\centerline{\huge \bf Conformal Fluctuations of the Metric \normalsize
\footnote{Partially supported by ESO, Praxis, Sapiens, CERN and FCT, grant numbers  
PESO /P/PRO/15127/1999 and CERN/P/FIS/40119 /2000}}
\vs

\vs

\centerline{Alex H. Blin}
\vs

\centerline{Centro de F\ii sica Te\'orica da Universidade, P-3004-516 Coimbra, Portugal}

\centerline{\texttt{alex@teor.fis.uc.pt}}
\vs

\vs

\ni {\sl Abstract} - At the level of the Planck scale, the spacetime metric has to be considered a quantum variable. Conformal quantum fluctuations of the metric tensor are studied here. They lead to an extra term in the Einstein equations which can be identified with a cosmological constant. Quantum fluctuations may therefore contribute to an accelerated expansion of the universe, in accordance with newer observational data.

\section{Introduction}

The cosmological constant was originally introduced by Einstein to allow for a static solution of the gravitational equations describing the universe, a requirement which turned out later to be in contradiction with observation. Einstein's equations with the cosmological constant $\Lambda$ read
\be
R_{\mu\nu}-\iv{2}g_{\mu\nu}R+\Lambda g_{\mu\nu}=-\fr{8\pi G}{c^4}T_{\mu\nu}\ .
\ee
Nowadays $\Lambda$ is getting revived interest again in the context of inflationary scenarios, string theory and particle physics, and of the actual expansion rate of the universe. It is the latter which is addressed in the present work. Recent observations of type Ia supernovae give ever stronger evidence of an expansion of the universe which is currently accelarating 
(e.g. \cite{perl,riess}), contrary to what would be expected when considering the gravitational attraction of matter existing in the universe. The acceleration may explain also the exsistent discrepancies between the age of the universe and age estimates of stars in globular clusters, although the latter are still under dispute \cite{chaboyer,lebreton}.
\vs

Combining considerations of General Relativity and Quantum Mechanics leads to the concept of a limiting scale, the Planck scale. This scale is set by the energy of a particle whose Schwarzschild radius is comparable to its Compton wavelength. The Planck energy is
\be
E_P=c^2\sqrt{\fr{\hbar c}{2G}}\ .
\ee
corresponding to the Planck length
\be
l_P=\sqrt{2G\hbar c}/c^2\cong 10^{-35}m \ .
\ee
The fact that there exists a minimum length scale at which General Relativity and Quantum Mechanics are fundamentally interwoven indicates that the metric tensor  $g_{\mu\nu}$ itself has to be regarded a quantum variable.

\section{Conformal fluctuations}

The simplest quantum variations of the metric are conformal ones. They are certainly not the most general ones. Conformal changes of the metric have however the advantage to keep the light cone structure of spacetime intact, which is important for not violating causality at any instant of the fluctuation. (The fluctuations lead however to a Planck size fuzziness of the light cone {\sl after} averaging over the fluctuations). This type of fluctuations is written as
\be
g_{\mu\nu}=\Phi^2\bar{g}_{\mu\nu}=(1+\varphi)^2\bar{g}_{\mu\nu}\hs .
\ee
where $\bar{g}_{\mu\nu}$ denotes the "classical" or "background" metric about which the fluctuations occur. The Einstein equations can be derived from variation of the action $(\hbar=c=1)$
\be
S=S_g+S_m=\iv{16\pi G}\int \rmd^4x \sqrt{-g}R +S_m\ 
\ee
where $S_g$ is the gravitational (Hilbert) action and $S_m$ the matter part. Conformal fluctuations change the scalar curvature to
\be
R=\fr{\bar{R}}{(1+\varphi)^2}+\fr{6\bar{g}^{\mu\nu}\partial_{\mu}\partial_{\nu}
\varphi}{(1+\varphi)^3}\hs .
\ee
where the overbar indicates the classical value, i.e. the one obtained when no fluctuations are present. In the special case of flat background spacetime the gravitational action reduces to
\be
S_g=-\iv{8\pi^2\lambda^2}\int\rmd^4x \partial^i\varphi\partial_i\varphi\ , 
\ee
with the quantity $\lambda$ related to the Planck length via
\be 
\lambda^2=\fr{l_P^2}{3\pi}\ .
\ee
The "wrong sign" of the action does not imply that the action is unbounded from below. The apparent unboundedness is an artifact of not taking the correct functional measure when passing to Euclidean space \cite{mazur,dasgupta}. It also means that $\varphi$ is not a freely propagating field, but it does have a vacuum expectation value. 
The proper distance acquires a lower limit of size $\lambda$. The fluctuation average of the squared four-distance becomes 
\be
<x^2>=x^2+\lambda^2\ .
\ee
Thus spacetime is coarse grained at the level of the Planck scale. As a consequence, divergences related to Green's functions (propagators) disappear, without the need for regularization or renormalization \cite{blin}.

\section {Variation of the action with fluctuations}

The variation of the action includes the variation of the background metric and of the fluctuations:
\be
\delta S=\fr{\delta S}{\delta g^{ik}}\delta g^{ik}=\fr{\delta S}{\delta \bar{g}^{ik}}\delta\bar{g}^{ik}+\fr{\delta S}{\delta\varphi}\delta\varphi\ .
\ee
For the matter part one has
\be
\delta S_m=-\iv{2}\int \rmd^4x\sqrt{-g}\delta g_{mn}T^{mn}\ .
\ee
Let me use Friedmann-Weyl cosmology to describe the most symmetric type of universe.  The line element is
\be
\rmd s^2=\rmd t^2+g_{kl}\rmd x^k\rmd x^l
\ee
and the cosmological principle of homogeneity and isotropy leads to the Robertson-Walker metric
\be
\rmd s^2=\rmd t^2-Q^2(t)[\fr{\rmd r^2}{1-kr^2}+r^2(\rmd\theta^2+\sin^2\theta \rmd\phi)]\ .
\ee
The quantity $Q(t)$ describes the evolution of the size of the universe and the sign of $k$ determines whether the universe is classically open or closed. The Einstein equations can be written as
\be
\bar{R}_{ik}-\iv{2}\bar{g}_{ik}\bar{R}+\bar{g}_{ik}\Lambda=-8\pi G T_{ik}
\ee
where a cosmological constant arises of the form 
\be
\Lambda=-\iv{4}(8\pi G \bar{g}^{mn}T_{mn}-\bar{R})\ .
\label{Lambda}
\ee
To complete the derivation of the Einstein equations, an assumption on the matter distribution in the universe is needed. The simplest and most symmetrical distribution considers homogeneous comoving matter (dust approximation):
\be
T_{kl}=\rho u_k u_l\hs \mbox{with\ } \{u_k\}=(1,0,0,0)\ .
\ee
Energy conservation $T^{kl}_{;l}=0$ leads to
\be
\rho(t)=\fr{\rho_0 Q_0^3}{Q^3}\ .
\ee
The variation of the action yields finally the equation of motion 
\be
\ddot{Q}Q^2-\dot{Q}^2Q-kQ+4\pi G \rho_0 Q_0^3=0
\ee
with the solution
\be
t-t_0=A \int_1^{Q/Q_0}\rmd q \sqrt{\fr{q}{B q^3-C q+D}}
\ee
where $A,B,C,D$ are known functions of $\rho_0,H_0,k,Q_0$.

\section{Numerical results and discussion}

Figure 1 shows the time evolution of the expansion parameter $Q$ (full line) in units of the present epoch value $Q_0$, assuming a present time density of $0.3$ in units of the critical density and a deceleration parameter
\be
d_0=-\fr{\ddot{Q}_0Q_0}{\dot{Q}^2_0}
\ee
of value -0.07 (the negative sign meaning accelerating expansion). This results in a cosmological constant
\be
\Lambda=4\pi G \rho_0-3H_0^2d_0
\ee
with the value 0.66 in units of the squared Hubbel constant, in agreement with recent observational data \cite{stompor,bernardis}. The same figure shows also the classical $\Lambda=0$ time evolution (dashed line), assuming the same present epoch density. The corresponding expansion is of course decelerating in that case (with the deceleration parameter +0.15). The time axis is drawn in units of the inverse Hubble constant and the present epoch time is taken to be zero. It should be noted that due to the dust approximation the results are not reliable for times in the remote past. The two curves show nevertheless that the quantum fluctuations result in an increased age of the universe. The future expansion rate is visibly influenced by the repulsive effect of the fluctuations.
\vs

%\vspace{8cm}
%
%\hspace{1.5cm}\special{eps:expand.eps x=12cm y=8cm}
%\vs
%
%\centerline{Figure 1 - $Q$ vs. $t$ in relative units (see text), with (full line) and}
%\centerline{without (dashed) quantum fluctuations.}
%\vs

\begin{figure}[h]
\centerline{\epsfig{file=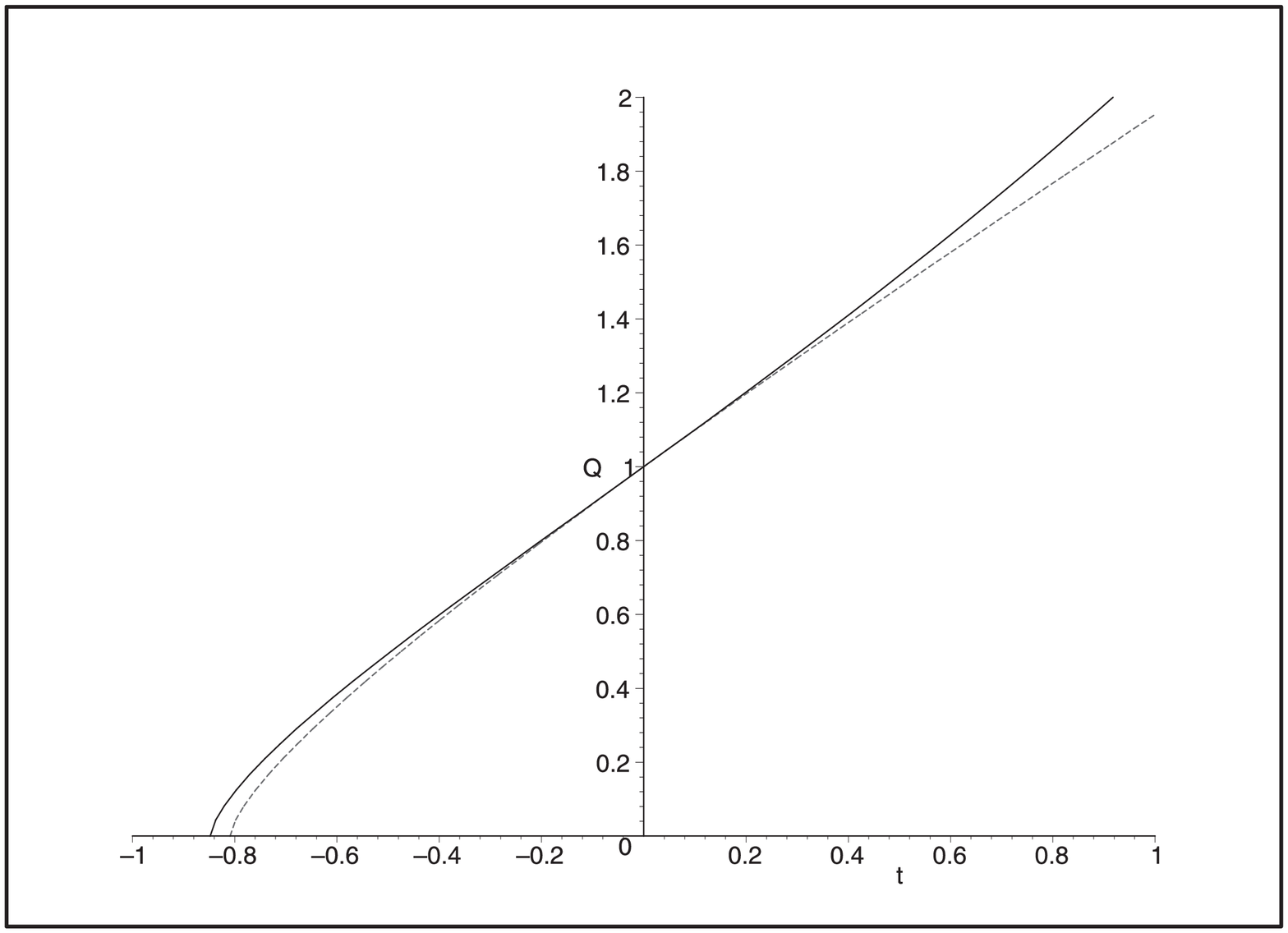,height=8cm,width=12cm}}
%\vspace{10pt}
%\caption{$Q$ vs. $t$ in relative units (see text), with (full line) and without (dashed) quantum fluctuations.}
%\label{fig1}
\end{figure}
\centerline{Figure 1 - $Q$ vs. $t$ in relative units (see text), with (full line) and}
\centerline{without (dashed) quantum fluctuations.}
\vs

In summary, conformal fluctuations of the metric are shown to contribute to a cosmological constant. This may explain why an accelerated expansion of the universe is observed. It might seem surprising that fluctuations at the level of the Planck size have consequences on the expansion of the universe as a whole. One should keep in mind, however, that the fluctuations exist in any and every point of space, the contributions adding up to a macroscopic effect.
\vs

As mentoned before, degrees of freedom other than the conformal one may have to be considered to reflect more fully the effects of quantum gravity. Due to this and to the dust approximation employed in the present calculations, together with our incomplete knowledge of the present day parameters, the results are to be considered qualitative rather than quantitative. Nevertheless it seems clear that quantum fluctuations have to be taken seriously as they affect macroscopic observables.

\end{document}